\documentclass[aps,prl,twocolumn,superscriptaddress,10pt,showpacs]{revtex4}
\usepackage[dvips]{graphics}
\usepackage{graphicx}
\usepackage{amsfonts}
\usepackage{amssymb}
\usepackage{amsmath}
\usepackage{subfigure}
\usepackage{wasysym}
\DeclareMathAlphabet{\mathpzc}{OT1}{pzc}{m}{it}

\begin{document}

\title{Stable structures with high topological charge in
nonlinear photonic quasicrystals}
\author{K.\ J.\ H.\ Law}
\affiliation{Warwick Mathematics Institute, University of Warwick,
  Coventry CV4 7AL, UK}
\affiliation{Department of Mathematics and Statistics,
University of Massachusetts,
Amherst MA 01003-4515, USA}
\author{Avadh Saxena}
\affiliation{Theoretical Division and Center for Nonlinear Studies,
Los Alamos National Laboratory, Los Alamos, NM 87545, USA}
\author{P.\ G.\ Kevrekidis}
\affiliation{Department of Mathematics and Statistics,
University of Massachusetts,
Amherst MA 01003-4515, USA}
\author{A.\ R.\ Bishop}
\affiliation{Theoretical Division and Center for Nonlinear Studies,
Los Alamos National Laboratory, Los Alamos, NM 87545, USA}


\begin{abstract}
Stable vortices with topological charge of $3$ and $4$ are
examined numerically and analytically in photonic quasicrystals
created by interference of $5$ as well as $8$ beams, for cubic as well as saturable nonlinearities .
Direct numerical simulations corroborate
the analytical and numerical linear stability analysis predictions
for such experimentally realizable structures.
\end{abstract}

\maketitle

{\it Introduction}. The study of vortices
has been a principal theme of interest in dispersive
nonlinear systems with applications including, among
others, Bose-Einstein condensates (BEC), and
nonlinear optical media \cite{pismen,fetter,kivshar_bk}.  More
recently, such states have been studied in
settings with some discrete spatial symmetry i.e., nonlinear lattices.
There, the notion of ``discrete vortices'' \cite{mk01}
arose and was subsequently intensely studied in both discrete and
quasi-continuum media; see e.g. \cite{led,pgk} for relevant reviews.
This led to the experimental realization of unit-charge ($S=1$)
coherent structures in saturably nonlinear photorefractive media
[such as SBN:75(Sr$_0.75$Ba$_0.25$Nb$_2$O$_6$)] in \cite{dnc,ysk},
and the exploration of higher charge ($S=2$) ones in square
and hexagonal/honeycomb lattice settings \cite{us2}.
A multipole soliton necklace of out-of-phase neighboring lobes
in a square lattice was identified experimentally and theoretically
in \cite{yang1} from initial condition of a wide $S=4$ gaussian beam.

While regular lattices have been mostly
studied \cite{led}, more recently experimental developments
have enabled the study of photonic
quasicrystals in photorefractive media
\cite{segev_nature}, and have spurred a correspondingly intense
theoretical activity \cite{ablowitz}.  We also
note that recent experiments have emerged on non-square
optical lattices for ultracold atoms in the BEC case \cite{sengstock}.
It is then natural to expect that quasi-crystals are well within
experimental reach in this regard, as well.

Motivated by these developments, we illustrate the unique
ability of such lattices (with saturable or cubic
nonlinearity) to sustain stable vortices of higher topological charge,
such as $S=3$ and $S=4$. Direct numerical simulations
reveal the robustness of such
modes. On the other hand,
perhaps counter-intuitively (but as can be analytically predicted),
lower charge vortices are found to be unstable,
and this instability
is also dynamically
monitored. 


{\it Theoretical Setup}.
We 
introduce the following
non-dimensionalized evolution equation:

\begin{equation}
\left[i\partial_z + \frac{1}{2}\nabla^2 + F(|U|^2)-V({
\bf x})\right]U=0.
\label{eq1}
\end{equation}

The (saturable) photorefractive nonlinearity is $F(|U|^2)=-1/(1+|U|^2)+1$,
where $U$ is the slowly varying amplitude
of a probe beam normalized by the dark irradiance of the crystal $I_d$
\cite{efrem, kivshar_bk},
and $V$ represents modulation of the refractive index from interfering
linearly propagating waves normal to the probe beam.
In a Kerr medium the nonlinearity reads
$F(|U|^2)=|U|^2$, and this case also includes the interpretation of $U$ as
a mean-field wavefunction of an atomic Bose-Einstein condensate
\cite{stringari}, while the potential $V$ is either modulation of the
refractive index in the former case or an externally applied
field in the latter.

\begin{figure}[tbh]
\begin{center}
\includegraphics[width=0.45\textwidth]{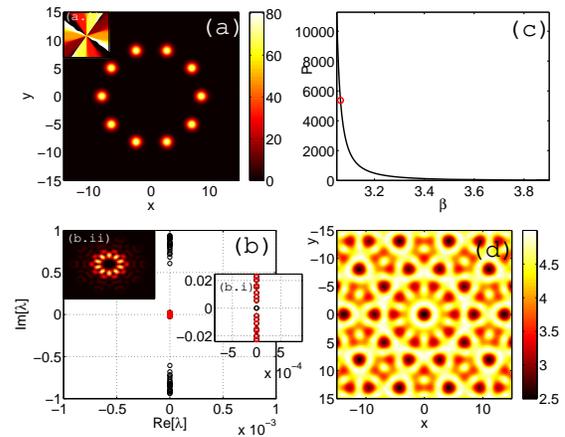}
\end{center}
\caption{(Color online) The stable $S=4$ vortex in a quasi-crystal
lattice of $N=5$ and with a saturable nonlinearity.
The profile and phase are depicted in panels [a(.i)],
the linear spectrum in panel (b), Fourier spectrum
in the inset panel (b.i), and continuation of the power,
$P=\int |U|^2 d{\mathbf x}$, as a function of the
propagation constant, $\beta$, in panel (c).
The $N=5$ lattice is depicted in (d).}
\label{fig1}
\end{figure}

The potential $V$ is taken to be of the form $E/(1+I(x))$, where
$I(\textbf{x})=
\frac{1}{N^2}\left| \sum_{j=1}^N e^{ik\textbf{b}_j \cdot \textbf{x}}\right|^2.$
In the photorefractive paradigm,
this is the optical lattice intensity function formed by $N$ interfering
beams in the principal directions $\textbf{b}_j$ with
periodicity $2\pi/k$.  We will consider the cases of $N=5$ and $N=8$.
Here 1 is the lattice peak intensity, $z$ is the propagation
constant, $\textbf{x}=(x, y)$ are transverse distances, $k=2\pi/5$ is the
wavenumber of the lattice, and $E=5$ is proportional to the
external voltage.  Recently, such a setting has been explored theoretically
for positive lattice solitons \cite{ablowitz,weinstein},
but we extend the considerations here to vortex
solutions.

\begin{figure}[tbh]
\begin{center}
\includegraphics[width=0.45\textwidth]{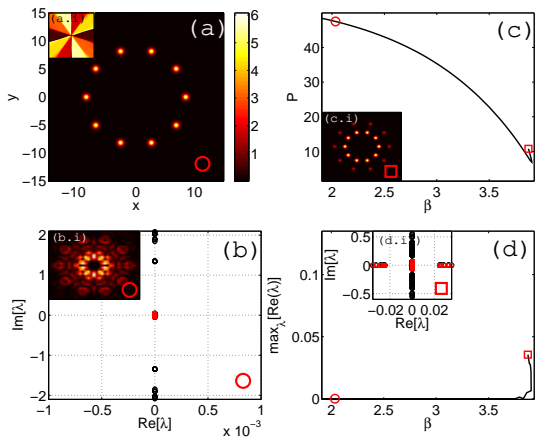}
\end{center}
\caption{(Color online) Panels (a-c) are the same as Fig. \ref{fig1}
except for a cubic nonlinearity.  Panel (d) shows the growth rate, or
${\rm max}_{\lambda} [Re(\lambda)]$. The insets, (c.i) and (d.i) depict
the profile and linear spectra, respectively, of the highly unstable
solution indicated by a red square on the branches in (c,d), which
collides with the main branch and disappears in a saddle node bifurcation
close to the phonon band edge.}
\label{fig2}
\end{figure}

The possible charge, $S$, of vortices (the number of $2 \pi$
phase shifts across a discrete contour comprising the solution)
is bounded by the symmetry of the
lattice \cite{kartashov}. A lattice with $n$-fold symmetry has
natural contours of $2n$ sites.  Hence,
taking into account the degeneracy of vortex
anti-vortex pairs $\{S,-S\}$, one has
$0 \leq S \leq n$, with the cases of $S=0,n$ being the trivial flux cases
of in-phase and out-of-phase neighboring lobes, respectively.
The quasi-crystal 
with $N=5$ has $n=5$, while for $N=8$,
$n=4$.  Hence, the highest possible charge, $S=n-1$,
is $S=4$ for the case of $N=5$ and $S=3$ for $N=8$.

Considering the quasi-one-dimensional contour
of excited sites (depending on the
respective amplitudes of the lattice and the probe field),
and within the context of coupled mode theory \cite{discrete_mi} in which the
probe field is expanded in Wannier functions \cite{wannier},
one can obtain insights about the stability of the vortices
within the framework of a discrete Nonlinear Schr{\"o}dinger
equation \cite{pgk},
$i \dot u_n = -\varepsilon(u_{n+1}+u_{n-1}-2 u_n)-|u_n|^2 u_n$.
In that context and based on either the
modulational instability of \cite{discrete_mi},
or through empirical numerical testing \cite{kartashov}
or, more rigorously, via Lyapunov-Schmidt perturbative expansions
around the so-called {\it anti-continuum} (AC) limit of zero coupling
($\varepsilon=0$) \cite{peli},
it is known that lobes which are phase-separated
by greater than $\pi/2$ are stable next to each other, while those
separated by less than $\pi/2$ are unstable.
A simple intuitive
argument for this situation is that the effective potential
which out-of-phase neighboring nodes exert on one another through the
focusing non-linearity is repulsive, and,
hence, remain localized in their respective separate wells.  On the other
hand, if the neighbors are in-phase, then the effective
neighboring potentials are attractive 
and hence the solution is unstable to remaining localized in separate
wells.
The possible relative phases interpolate between these cases,
with $\pi/2$ being exactly in the middle.
A similar discussion is used in \cite{kivshar_ref} in order
to justify (upon suitable phase variation) the existence of
soliton clusters in bulk media.
This leads to stability of the {\it higher} charged vortices
for contours of larger numbers of nodes (see also \cite{us2}).
We briefly review the
Lyapunov-Schmidt argument.
In the limit $\varepsilon \rightarrow 0$ one can construct exact
solutions of the form
$u_{j}=\sqrt{\mu} e^{\{- i (\beta t- \theta_j) \}}$
for any arbitrary $\theta_{j} \in [0,2\pi)$ \cite{peli}.
The case we are considering is that of $\theta_j = jS\pi/n$.
We linearize
around the solution for $\varepsilon=0$
and the condition for existence
of solutions with $\varepsilon>0$
reduces to the vanishing of a vector function
${\bf g}({\boldsymbol \theta})$ of the
phase vector 
${\boldsymbol \theta}=(\theta_1, \dots, \theta_N)$,
where

\begin{equation}
g_j \equiv \sin(\theta_{j-1}-\theta_{j})+\sin(\theta_{j+1}-\theta_{j}),
\label{eq3}
\end{equation}

\noindent subject to periodic boundary conditions.
This includes the discrete reduction of the vortex
solutions for $0 \leq S \leq n$ above.
The fundamental contours $M$
will have length $|M|=2n$, and
$|\phi_{j+1}-\phi_j|=\Delta \phi=\pi S/n$
is constant for all $j \in M$,
$|\theta_1-\theta_{|M|}|=\Delta \theta$ and
$\Delta \theta |M| = 0$ ${\rm mod}$ $2\pi$.

For the contour M, there are $|M|$
eigenvalues $\gamma_j$ of the $|M| \times |M|$
Jacobian ${\cal M}_{jk}=\partial g_j/\partial \theta_k$ of
the diffeomorphism given in Eq. (\ref{eq3}).
The eigenvalues of this matrix $\gamma_j$ can be mapped 
eigenvalues of
the full linearization.  
In particular, eigenvalues of the linearization,
denoted $\lambda_j$, are given to leading order
by the relation \cite{peli}
$\lambda_j = \pm \sqrt{2 \gamma_j \varepsilon}$.
Thus, solutions are stable to leading order if $\gamma_j<0$
(so $\lambda_j \in i \mathbb{R}$) and unstable
if $\gamma_j>0$ (so $\lambda_j \in \mathbb{R}$).
We have $\gamma_j = 4 \cos\left(\Delta \phi\right)
\sin^2\left(\frac{\pi j}{|M|}\right)$
and so these cases correspond exactly to 
$\Delta\phi>\pi/2$ (or $S>n/2$) and
$\Delta\phi<\pi/2$ (or $S<n/2$).
In the boundary case of $\Delta\phi=\pi/2$,
one needs to expand to the next order in the
Lyapunov-Schmidt reduction.  We note that a so-called staggering
transformation along the contour, $u^d_j=(-1)^j u^f_j$
allows the above conclusions for the focusing problem 
to be mapped immediately to the defocusing problem (with a
change in the sign of the nonlinearity).  We
do not consider the defocusing case further here.
The above considerations illustrate the
expectation that $S=3$ vortices may be
stable in the $N=5$ and $N=8$ cases, and the $S=4$ vortex may
be stable in the $N=5$ case.

{\it Numerical Results}. We now turn to numerical computations.  
We also explore
the evolution of different $S$ radial gaussian beams.

\begin{figure}[tbh!]
\begin{center}
\includegraphics[width=0.45\textwidth]{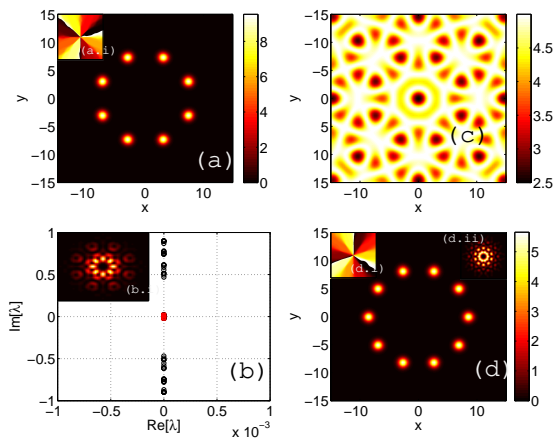}
\end{center}
\caption{(Color online) Panels (a,b) are the same as the previous figures
for the stable $S=3$ vortex in the $N=8$ quasi-crystal lattice (c).
Panel (d) is the $S=3$ vortex for $N=5$
and [d.(i,ii)] are the phase and
Fourier spectrum, respectively, of this solution.
For both solutions, $\beta=3.4$.}
\label{fig3}
\end{figure}

First, we confirm the expectation of stability of the
$S=4$ vortex for saturable and cubic
nonlinearities, over continuations in the semi-infinite gap
(see Figs. \ref{fig1} and \ref{fig2}, respectively).  The profiles and phases
are depicted in panels [a(.i)], linear spectra in panels (b), Fourier spectra
in the inset panels (b.i), and continuations of the power,
$P=\int |U|^2 d{\mathbf x}$,
as a function of the propagation constant, $\beta$, in panels (c).
The power of the solution branches differs
substantially between nonlinearities,
and the power of the branch of
saturable solutions approaches some resonant frequency at which
$dP/d\beta \rightarrow \infty$ and $P \rightarrow \infty$
(see Fig. \ref{fig1} (c)).
The lattice is depicted in Fig. \ref{fig1} (d), while
Fig. \ref{fig2} (d) shows the maximal perturbation growth rate, or
${\rm max}_{\lambda} [Re(\lambda)]$, corresponding to the branches
in Fig. \ref{fig2} (c).

\begin{figure}[tbh]
\begin{center}
\includegraphics[width=0.45\textwidth]{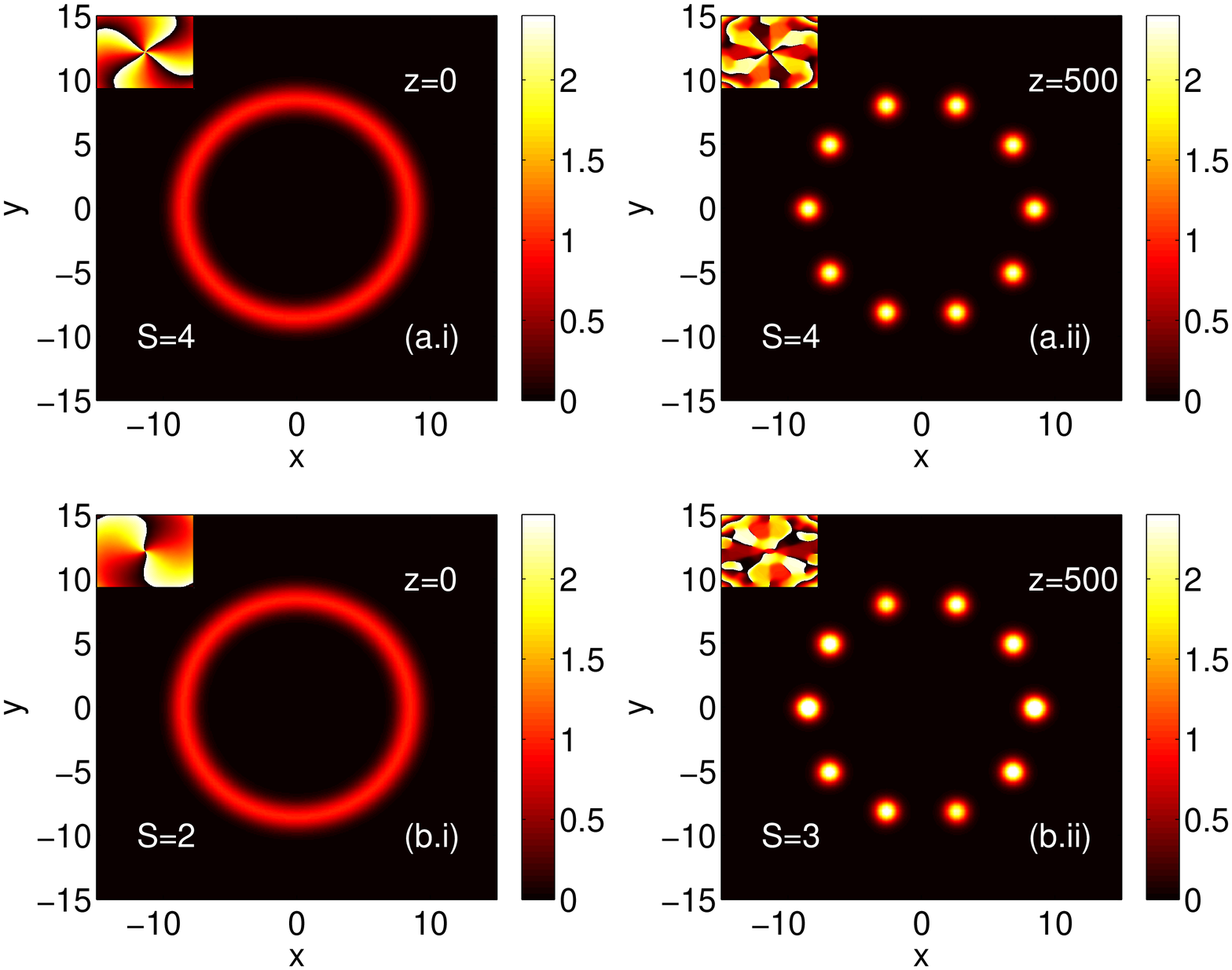}
\end{center}
\begin{center}
\includegraphics[width=0.45\textwidth]{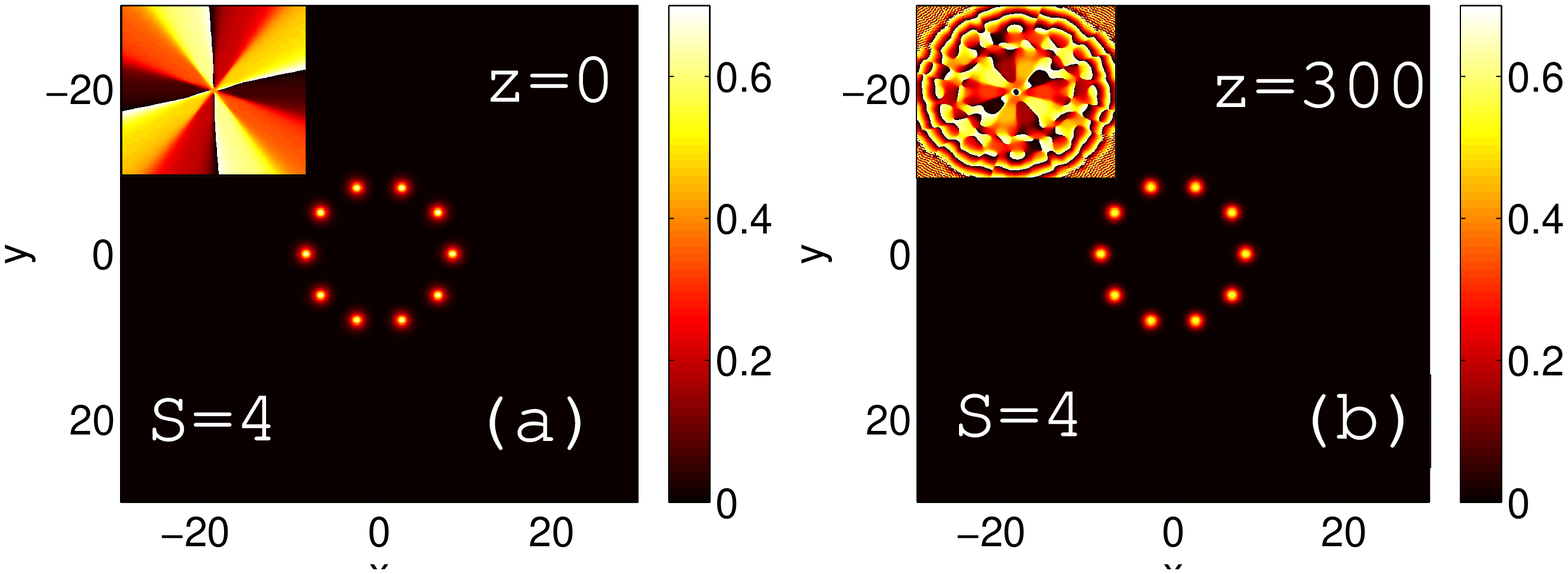}
\end{center}
\caption{(Color online) Initial conditions (a.i,b.i)
and profiles at a later time (a.ii,b.ii) of the $S=4$ and $S=2$
radial Gaussian initial
conditions for a saturable nonlinearity, 
with a ``tight dissipation layer'' (see text), $D=16$.
The bottom is similar to the top but for $D=24$.}
\label{fig4}
\end{figure}


For the structures we consider, there
is one pair of eigenvalues at the origin accounting for
the U$(1)$ (phase)  invariance and the other $2n-1$ eigenvalue pairs
associated to the excited lobes all have negative energy,
hence being candidates for instability \cite{meer}, and are all
either purely imaginary or purely real.  If real, the
instability is immediate, while if imaginary, instability may still
arise due to their collision
with the phonon band,  resulting in a
Hamiltonian-Hopf bifurcation and eigenvalue quartets.
The spectral plane, with the negative
energy modes indicated by red squares, for the saturable and cubic cases
are given in panels (b) of Figs. \ref{fig1} and \ref{fig2}, respectively.
Panel (b.ii) in Fig. \ref{fig1} is a closeup of the origin showing the 9
negative energy pairs
close to the origin and the one pair at the origin.  The potential instability
arising from these negative energy modes is prevented by their proximity to the
origin, and distance from the phonon band.  The expected saddle-node
bifurcation \cite{yang,ourdef} occurs close to the band edge (which we
computed as $\approx 3.9$) in which the main solution collides
(and disappears) with
an unstable solution branch of a configuration with additional populated sites
external (and in phase) to the original contour.

\begin{figure}[tbh]
\begin{center}
\includegraphics[width=0.45\textwidth]{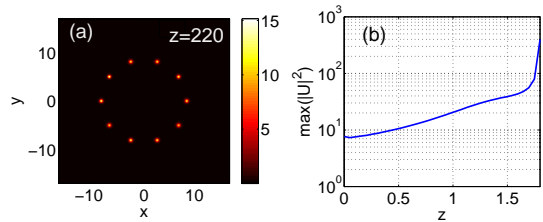}
\end{center}
\caption{(Color online) The dynamics of the unstable $S=2$ vortex in the
case of a cubic nonlinearity.  Evolution of the same solution
with the same perturbation of random noise
with $5\%$ of the initial maximum amplitude of the field can lead to
robust structures that persist for long distances (a) and almost
immediate collapse in different trials (b).}
\label{fig5}
\end{figure}

Next, we present results of the $S=3$ vortex in both the $N=8$
(Fig. \ref{fig3}(a,b)) and $N=5$ (Fig. \ref{fig3} (d)) cases for $\beta=3.4$.
Panel (c) depicts the $N=8$ lattice, and [d.(i,ii)] are the phase and
Fourier spectrum, respectively, of the solution in (d).
These solutions are both stable,
and again there is a resonance in the
semi-infinite gap (not shown) similar to what was seen in Fig. \ref{fig1}.
The vortices for $S<3$ are unstable (not shown).

To examine the potential experimental realizability of the
above waveforms,
we launch a radial Gaussian beam with topological charge $S=4$ of the form
$e^{iS \theta-(r-R)^2/(2b^2)}$ with $(r,\theta)$ denoting polar
coordinates, $R=8.5$,
approximately the radius of the contour, and $b=1$,
as an initial condition into the system with saturable nonlinearity
and monitor the evolution.
After a transient period, the configuration indeed settles into an
$S=4$ vortex contour. However, notice that this simulation has been
performed with a ``tight dissipation layer'' 
i.e. using an extra term $-i\Gamma$ on the right-hand side of Eq. \ref{eq1}) of
$\Gamma=1-{\rm tanh}(D-{\bf r})$ with $D=16$. This initially 
absorbs the shed radiation, and subsequently affects very little the 
intensity distribution. However, the phase dynamics may be 
sensitive to the presence of such a layer:
imposing a wider such dissipation layer with $D=24$,  the solution actually 
never settles into one of constant charge; this topological
instability
effect has been analyzed e.g. in
\cite{kiv_ref4,kiv_ref3,kiv_ref1,kiv_ref2}. 
Specifically, vortices may nucleate in the 
very small amplitude region and pass in and out of the main 
configuration (without affecting its intensity). 
Note that the above suggests that such effects could 
be avoided experimentally if some form of dissipation is imposed. 
For comparison, we launch a similar initial
condition with $S=2$ and notice that it never settles into a
stable configuration of fixed charge {\it independently of the 
dissipation layer size}
(and despite its seemingly robust intensity distribution).

Figure \ref{fig4} (top two panels) present the initial conditions (a.i,b.i)
and profiles for a long evolution (a.ii,b.ii) of the $S=4$ and $S=2$ initial
conditions, respectively, for saturable nonlinearity and $D=16$.
The charge of each fluctuates, as power is shed and
vortices nucleate in the surrounding low amplitude regions
and enter and leave the contour as the solution traces a stationary state.
However, for the $S=4$ initial condition, the field settles into
a solution of constant charge 4 for $D=16$, while
for the $S=2$ initial condition, the phase continues to fluctuate
throughout the numerical experiment.
These results are typical in this setting.  
For $D$ large (bottom panel of Fig. \ref{fig4}) the charge may never
settle (topological phase instability). However, this does not
contradict the linear stability results (which we have confirmed separately
for near stationary configurations).  The intial condition
$e^{iS \theta-(r-R)^2/(2b^2)}\cos^2(5\theta)$ is far from
a stationary configuration and not sufficiently modulated to 
prevent contamination of the resulting 
state by radiation, although e.g.
$\sum_{k=1}^{10}e^{ikS\pi/5-(x-c_{xk})^2-(y-c_{yk})^2}$ with $(c_{xk},c_{yk})$ 
the center of one of the wells, is sufficiently localized, and with
the latter initial condition, we observe topological 
stability (indeed without the initial turbulent fluctuating regime) for $S=4$ (see Fig. \ref{fig4})
but not for $S=2$.
Finally, Fig. \ref{fig5} shows
the evolution of unstable ($S=2$) vortices in the
presence of a cubic nonlinearity.
The evolution depends sensitively
on the particular initial condition.  Using the initial
condition $u=U(1+X)$ with
$X \sim 0.05{\rm max}_{\bf x}[U({\bf x})]$ uniform over [0,1],
two different particular instances can lead to significantly different
dynamics.  Either the phase merely reshapes as for the 
saturable nonlinearity, but the structure persists  [see
Fig. \ref{fig5} (a)],  or the solution collapses almost immediately,
as seen
from the maximum amplitude of the field in Fig. \ref{fig5} (b).
For larger additive noise,
collapse seems more likely from several sample trials.
The relevant mechanism involves one of the solution lobes exceeding
a minimum collapse threshold, leading to an ``in lobe'' collapse.

{\it Conclusions and future directions}.
We have demonstrated numerically stable vortices of topological
charge $S=3$ in quasi-crystals with $n=4$ and $5$ directions of
symmetry and $S=4$ with $n=5$, in the cases of both cubic and
 saturable focusing nonlinearities.  The negative energy
modes for these configurations remain close to the origin in the
spectral plane,
preventing collision with the phonon band, and can be 
experimentally realizable in photonic
quasi-crystals in a photorefractive (or a Kerr) medium.  This
has additionally been demonstrated by simulation of the evolution of a
radial Gaussian beam into such robust vortex states.
This is a prime prospect for an immediate future
experimental direction related to the present work.


{\it Acknowledgments.}
KJHL acknowledges LANL and CNLS for 
hospitality. The work was supported by NSF and DoE.


\begin{thebibliography}{999}

\bibitem{pismen} L.M. Pismen,
{\it Vortices in Nonlinear Fields}, (Oxford University Press,
Oxford, 1999).

\bibitem{fetter} A.L. Fetter,
Rev. Mod. Phys. {\bf 81}, 647 (2009).


\bibitem{kivshar_bk} Y.S. Kivshar and G.P. Agrawal,
{\it Optical Solitons: From Fibers to Photonic Crystals},
(Academic Press, London, 2003).

\bibitem{mk01} B.A. Malomed and P.G. Kevrekidis,
Phys. Rev. E {\bf 64}, 026601 (2001).

\bibitem{led} F. Lederer {\it et al.}, Phys. Rep.
{\bf 463}, 1 (2008).

\bibitem{pgk} P.G. Kevrekidis, {\it The Discrete Nonlinear
Schr{\"o}dinger Equation}, (Springer-Verlag, Heidelberg, 2009).

\bibitem{ysk} D.N. Neshev {\it et al.}, Phys. Rev. Lett.
{\bf 92}, 123903 (2004).

\bibitem{dnc} J.W. Fleischer {\it et al.}, Phys. Rev. Lett.
{\bf 92}, 123904 (2004).

\bibitem{us2} K.J.H. Law {\it et al.}, Phys. Rev. A {\bf 80},
063817 (2009); B. Terhalle {\it et al.}, Phys. Rev. A {\bf 79}, 043821
(2009).

\bibitem{yang1} J. Yang {\it et al.}, Phys. Rev. Lett.
{\bf 94}, 113902 (2005).

\bibitem{segev_nature} B. Freedman {\it et al.},
Nature {\bf 440}, 1166 (2006).

\bibitem{ablowitz} M. J. Ablowitz {\it et al.}, Phys. Rev. E {\bf 74},
035601(R) (2006).

\bibitem{sengstock} C. Becker {\it et al.},
New J. Phys. {\bf 12}, 065025 (2010).




\bibitem{efrem} N.K. Efremidis {\it et al.},
Phys. Rev. E {\bf 66}, 46602 (2002).


\bibitem{stringari} L.P. Pitaevskii and S. Stringari,
{\it Bose-Einstein Condensation},
(Oxford University Press, Oxford, 2003).


\bibitem{weinstein} Y. Sivan, {\it et al.}, Phys. Rev. E {\bf 78},
046602 (2008).

\bibitem{kartashov} Y. V. Kartashov, {\it et al.}, Phys. Rev. Lett. {\bf 95},
123902 (2005).

\bibitem{discrete_mi} D.N. Christodoulides and R.I. Joseph, Opt. Lett. {\bf 13}, 794 (1988);
Yu.S.~Kivshar and M.~Peyrard, Phys.~Rev.~A {\bf 46}, 3198 (1992).

\bibitem{wannier} W. Kohn,  Phys. Rev. {\bf 115}, 809 (1959).

\bibitem{peli} D.E. Pelinovsky, P.G. Kevrekidis, and D.J. Frantzeskakis,
Physica D {\bf 212}, 1 (2005); {\it ibid.} 20 (2005).

\bibitem{kivshar_ref} A. S. Desyatnikov and Y. S. Kivshar,
Phys. Rev. Lett. {\bf 88}, 053901 (2002).

\bibitem{meer} J. C. van der Meer, Nonlinearity {\bf 3}, 1041 (1990).

\bibitem{yang} J. Wang and J. Yang,  Phys. Rev. A. {\bf 77}, 033834 (2008).

\bibitem{ourdef} H. Susanto {\it et al.},
Physica D {\bf 237}, 3123 (2008).

\bibitem{kiv_ref4} A.S. Desyatnikov, Yu.S. Kivshar and L. Torner,
Prog. Optics {\bf 47}, 291 (2005).

\bibitem{kiv_ref3} A. Bezryadina, {\it et. al.}, Opt. Exp. {\bf 14}, 8317 (2006).



\bibitem{kiv_ref1} B. Terhalle, {\it et. al.}, Phys. Rev. Lett. {\bf 101}, 013903 (2008).

\bibitem{kiv_ref2} B. Terhalle, {\it et. al.}, Opt. Lett. {\bf 35}, 604 (2010).



%
%




\end{thebibliography}
\end{document}